\documentclass[preprint,review,12pt]{elsarticle}
\usepackage{graphics}
\usepackage{color}
\usepackage[dvips]{epsfig}
\usepackage{amsmath}
\usepackage{amssymb}
\usepackage{amsthm}
\usepackage{longtable}
\usepackage{lineno}

\begin{document}
	
\begin{frontmatter}
\title{Magnetic field at nucleus for ionized atoms}
\author[a]{Andrzej Tucholski}
\ead{andrzej.tucholski@fuw.edu.pl}
\address[a]{University of Warsaw, Heavy Ion Laboratory, Pasteura 5a, 02-093 Warsaw, Poland}
\author[b]{Zygmunt Patyk}
\address[b]{National Centre for Nuclear Studies,  Pastera 5, 02-093 Warsaw, Poland}

\date{\today}

\begin{abstract}
The study of the magnetic properties of atoms and nuclei was performed. A linear dependence between the strength of the magnetic field at the nucleus and the effective charge divided by the major quantum number were analyzed and explained. The calculations of the magnetic field at the nucleus were done for a wide set of electron configurations within the framework of relativistic Dirac-Hartree-Fock method. Results are compared to experimental results of $g$-factor measurements. 
\end{abstract}


\end{frontmatter}
\pagebreak

\section{Introduction}
The time evolution of the quantum wave functions of the nucleus interacting with orbital electrons is very interesting by itself, but it also yields valuable insights into the structure of nuclei. Measurements of the magnetic properties of nuclei, especially magnetic moment measurements for higher spin states, enable the study of single-particle excitations in nuclei. The magnetic moment of the neutron differs significantly from that of the proton, enabling us to differentiate between single-particle excitations of neutron origin and those of the proton \cite{AS}. To measure higher spin states, we first have to measure magnetic moments for $2^+$ states as a reference, which allow us to measure strength of a magnetic field at nucleus. In the paper, we will demonstrate the universality and systematic nature of the magnetic properties of heavy ions.

The theory regarding the interaction between the magnetic moment of a nucleus and orbital electrons was established a long time ago, as described in references \cite{Goertzel} and \cite{Abragam}. This interaction causes periodic variations in the angular distribution of the emitted photons. The frequency of these variations is proportional to the product of the magnetic field generated by the electron shell and the magnetic moment of the nucleus.

\begin{figure}[h]
	\includegraphics[width=0.92\linewidth]{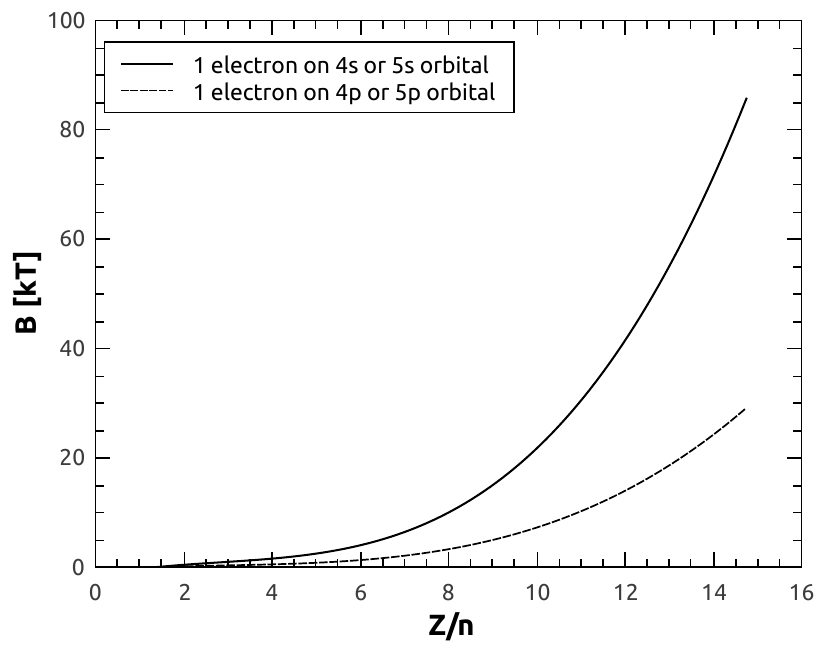}
	\caption{The magnetic field at the nucleus in an H-like ion, created by the orbitals 4s, 5s, and 4p, 5p, was calculated relativistically using the GRASP package.}
	\label{BZn}
\end{figure}

In case of H-like ions, for the orbital $ns$ (where $n$ is the major quantum number), the magnetic field is given by the non-relativistic equation $B = 16.685(Z/n)^3$~T, valid for small $Z$, where $Z$ is the charge of the nucleus \cite{Landau}. In case of the $p$ orbitals and higher ones, the Dirac-Hartree-Fock model must be applied. We used GRASP package developed by the groups at Oxford and Vanderbilt University, along with their collaborators. A comprehensive description of this computational framework can be found in Ref. \cite{Grasp,Grasp1,Grasp2,Grasp02}.

In Fig. \ref{BZn} we show the results of the calculations using GRASP code of the dependence of the magnetic field $B$ on $Z/n$. Due to the shape of the wave function, magnetic field of the electron on orbital $s$ is much stronger then that of the orbital $p$ and higher.

In case of heavy ions, this dependence becomes linear, and the charge $Z$ must be replaced by the effective charge $q_{eff}$ of the ion due to the screening effect. In our study, we compare the calculations to experimental results for ions with $Z$ in the range $44 < Z < 64$.

\section{Linear dependence of the magnetic strength}

The g-factor of nucleus cannot be directly measured during an experiment. Only the product $gB$ is measured, where $g$ is the $g$-factor and $B$ is the magnetic strength at the nucleus. The strength of the magnetic field depends on the configuration of electrons.

We performed an analysis of the set of possible electron configurations and magnetic field they generate. The number of ionized electrons depends on the velocity of ions traversing the solid matter (target). The specific configuration of electrons determines the total spin value $J$ of electrons and the magnetic field. According to the Bohr criterion, ions lose electrons whose velocities are less than those of the ions velocity traversing solid matter \cite{Bohr}. To get the number of ionized electrons, we based on two review papers analyzing more than 450 experimental measurements of the dependence of charge state versus velocity of ions \cite{Grande,Betz}. 
To calculate the magnetic field and spin values for different possible electron configurations, the GRASP code was used \cite{Grasp}. The results were compared with several experiments of g-factor measurements using Recoil in Vacuum (RIV) methods. 

The pioneering work of H.R. Andrews \cite{Andrews} in the field of hyperfine structure studies, introduces a significant contribution by highlighting the significance of the hard core and its correlation to the mean total spin value of electrons. This breakthrough study allowed for the measurement of the mean spin value of electrons by analyzing the asymptotic behavior of the attenuation factor G($t$) \cite{Andrews,Stuchbery_rev}.

The subsequent advancement in hyperfine structure studies resulted in the formulation of a semi-empirical formula to describe the dependence of the charge state on the velocity of ions. This finding enabled a significant reduction of the number of potential electron configurations for a given ion velocity in a particular medium.

We examined a series of experiments involving g-factor measurements for heavy ions with atomic numbers in the range of 44 to 64. The experiments show a linear dependence of the magnetic field versus effective charge divided by the major quantum number $n$.

In experiments, when the stripping of electrons is of the order of 30 electrons, like in \cite{Stone01}, the reconfiguration of electrons can be large. As a result, one can get different configurations leading to different magnetic fields at the nucleus \cite{Chen}. There is an unknown shape of the distribution for different configurations. One has to assume it. By doing that, the additional parameters are introduced to the fitting procedure. What is more, the unpaired electron in the configuration, changes dramatically the magnetic field strength at nucleus.

\begin{figure}[h]
	\includegraphics[width=0.92\linewidth]{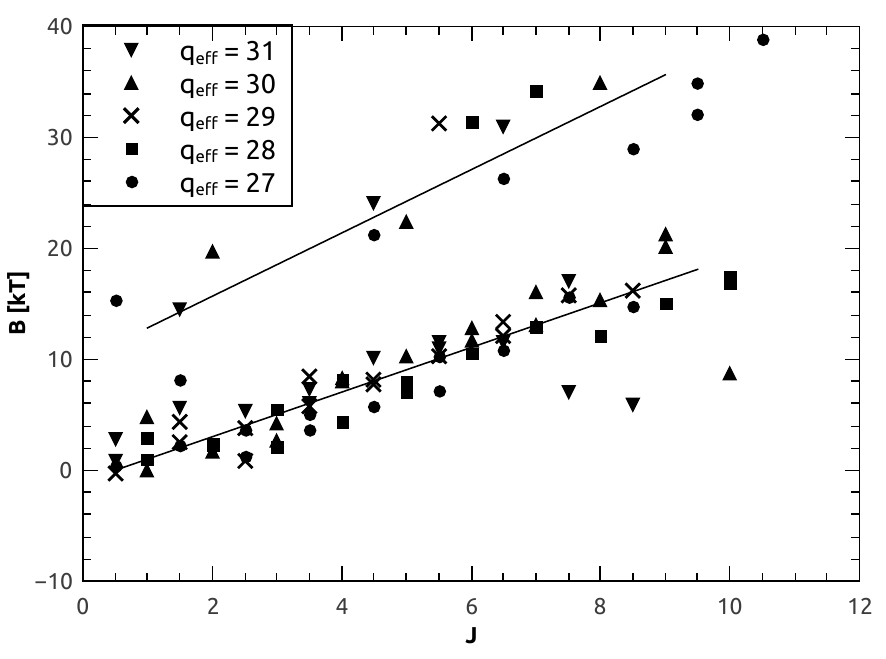}
	\caption{The magnetic field at the nucleus versus spin value for $^{132}$Te for the set of configurations: [Ar] + 3,4,5,6,7 electrons (effective charge $q_{eff}$ = 31,30,29,28,27). The upper line and the poits around it are for a set of configurations with one electron on the 4s orbital}.
	\label{132Te}
\end{figure}

We have made calculations with GRASP code for different configurations of electrons for a set of effective charges for several nuclei. In the case of $^{132}$Te \cite{Stone01}, we know from the experiment the mean ionization of the ions. We calculated the mean magnetic field for the set of different electron configurations and a set of different effective charges, with the GRASP code. (We kept the [Ar] electron configuration core unchanged and varied electrons in orbitals 3d and above). The results for $q_{eff}$ = 31,30,29,28,27 are shown in Fig. \ref{132Te}, where the magnetic field is plotted versus total electron spin.

\begin{figure}[h]
	\includegraphics[width=0.92\linewidth]{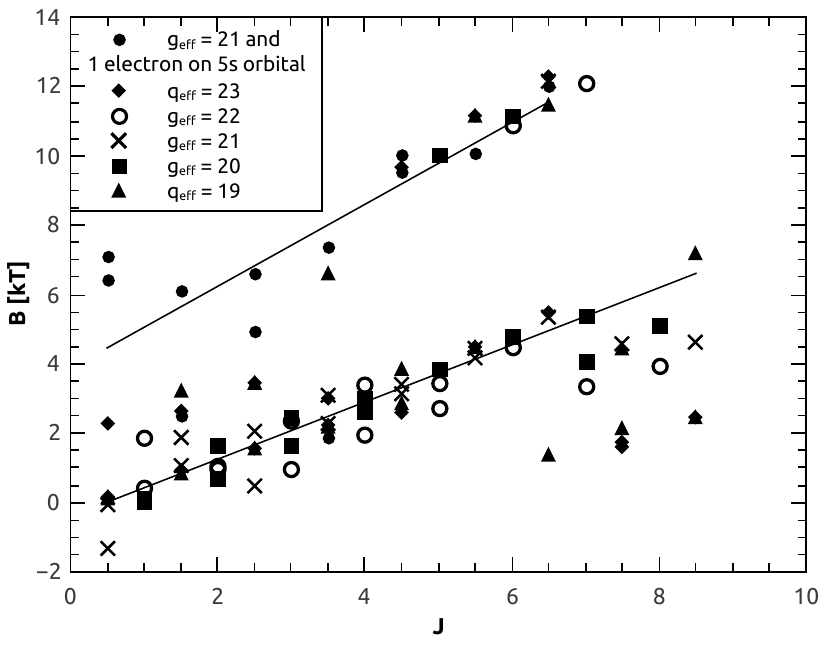}
	\caption{The magnetic field at the nucleus versus spin value for $^{150}$Sm for the set of configurations [Kr] + 3,4,5,6,7 electrons (effective charge $q_{eff}$ = 23,22,21,20,19). The upper line and the poits around it are for a set of configurations with one electron on the 5s orbital}
	\label{150Sm}
\end{figure}

One can see two lines and points around it. The upper one is for set of configurations such, that on the orbital 4s there is one electron. The lower one is for set of configurations such, that on the orbital 4s is pair of electrons. The mean value of B (calculated for the lower curve) is $<B> = 8.12~kT$, while the experimental value is $<B> = 8.8~kT$ \cite{Stone01}, and the mean value of spin is $<J> = 4.5 \hbar$ ($J = 4.5 \hbar$ in the experiment). Atoms during ionization prefer to keep electrons in pairs on orbitals if possible. That is why, in real systems, a pair of electrons is much more probable on orbital 4s then a single one, or it de-excites to the lower orbital d in a short time (as compared to precession time).

We repeated the procedure for $^{150}$Sm. The case is very interesting as compared to $^{132}$Te because this time the core (closed shells) is [Kr]. (We kept the [Kr] electron configuration core unchanged and varied electrons in orbitals 4d and above). The results for $q_{eff}$ = 23,22,21,20,19 are shown in Fig. \ref{150Sm}, where the magnetic field is plotted versus total electron spin.

\section{Comparison to experiment}

The mean value of the magnetic field is crucial in all methods of g-factor measurements. The magnetic field at the nucleus depends on the electron configuration and, consequently, on the effective charge of the atom. 
The effective charge accounts for the screening effect, where electrons partially shield each other, resulting in a reduced charge seen by the nucleus. The experimental values of effective charges versus velocity of the ions were summarized in \cite{Grande}.

\begin{figure} [h]
	\includegraphics[width=0.80\textwidth]{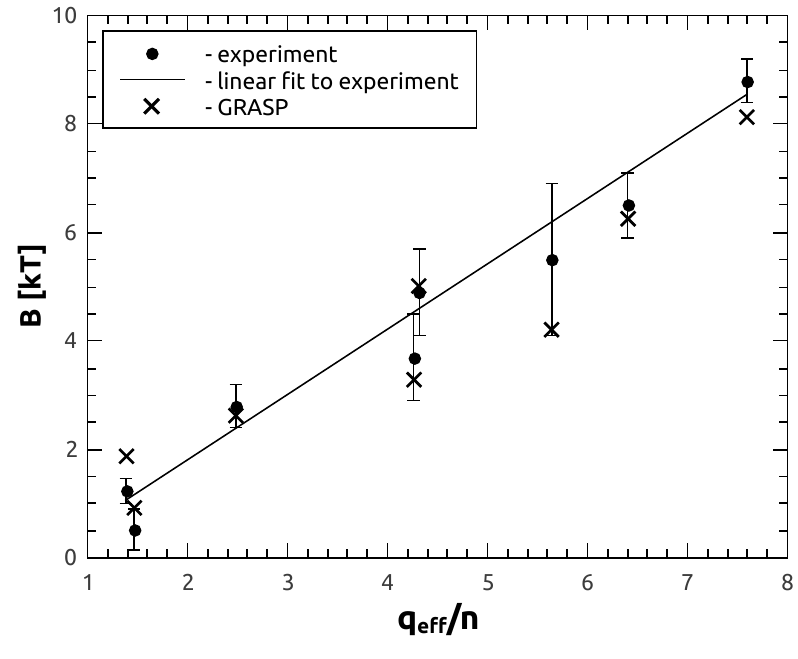} 
	\caption{The mean value of the magnetic fiend $B$ versus relative effective charge $q_{eff}/n$ for couple of experiments from Z=44 to Z=64, where $q_{eff}$ is taken from \cite{Grande}.}
	\label{comp_qeff}
\end{figure}

We took the experiments of Dafni et al. \cite{Dafni} who found $J=5\hbar$ for $^{144}$Gd and $J = 3\hbar$ for Fe ions, Andrews et al.
\cite{Andrews} reported $J=4.5\hbar$ for $^{110}$Cd ions, Stone et al. \cite{Stone01} who found $J=4.5\hbar$ for Te ions, Ward et al. \cite{Ward} found $J=4.5$ for $^{150}$Sm, Tucholski et al. \cite{AT} who found $J=4.5$ for $^{136}$Nd, and Allmond et al. \cite{Allmond} for several Sn isotopes. The results are accumulated in the Table \ref{table2}. In Fig. \ref{comp_qeff} we plotted the mean value of the magnetic field versus effective charge divided by principal quantum number. The ionic charge has been calculated from velocity using the Schwietz-Grande formula \cite{Grande} and later compared with that of Betz \cite{Betz}.

\begin{figure} [h]
	\includegraphics[width=0.80\textwidth]{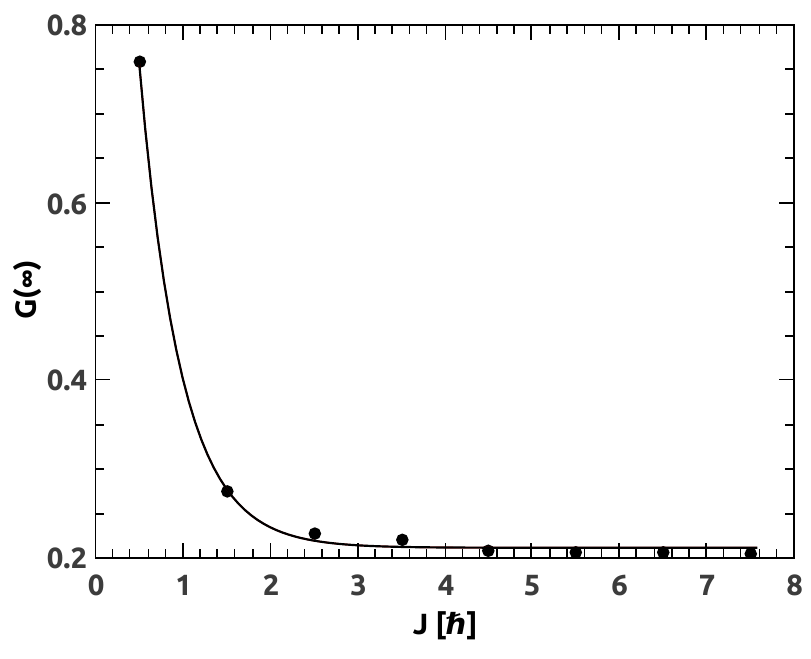} 
	\caption{The asymptotic behavior of the $G(x)$ versus spin $J$}
	\label{asymptoty}
\end{figure}

The question remains: why in almost all considered experiments, the spin is $J=9/2$ or close to that value (like $J=5$ in the Dafni experiment \cite{Dafni}). If we look into the asymptotic behavior of the attenuation factor $G(\infty)$ \cite{Stuchbery_rev} versus spin $J$ in Fig. \ref{asymptoty} (for $I=2$) it is independent of the magnetic field, and second, it is close to 1 for spin $J=1/2$ and much lower for higher spin values. We have calculated the mean spin value from the range of $1/2$ to $17/2$, weighted by the asymptotic value of G. The result is $J = 4.48$, which is close to $J = 9/2$. 

\begin{table}[h]
	\begin{center}
		\caption{The comparison of experimental values of the magnetic field with GRASP calculations.}
		\begin{tabular}{|p{1.1cm}|p{1.4cm}|c|p{1.2cm}|p{1.3cm}|c|c|p{1.7cm}|c|c|}
			\hline
			mean $\overline{q}_{eff}$ (Betz) & mean $\overline{q}_{eff}$ (Grande) & v/c & B [kT] exp. & B [kT] GRASP & experiment & Z & range of $q_{eff}$ & spin [$\hbar$] & $n$ \\\hline\hline
			2.91 & 8.3 & 0.01 & 1.23  & 1.44 & $^{136}$Nd \cite{AT} & 60 & [6$\cdots$10]  & J=9/2 & 6 \\\hline
			2.94 & 7.33 & 0.01 & 0.525  & 0.93 & $^{110}$Cd \cite{Andrews} & 48 & [5$\cdots$9] & J=9/2 & 5 \\\hline
			6.57 & 14.9 & 0.018 & 2.8  & 2.59 & $^{144}$Gd \cite{Dafni} & 64 & [13$\cdots$17] & J=5 & 6 \\\hline
			11.17 & 21.3 & 0.029 & 3.7  & 3.28 & $^{150}$Sm \cite{Ward} & 62 & [19$\cdots$23] & J=9/2 & 5 \\\hline
			13.3 & 21.6 & 0.037 & 4.9 & 5.01 & $^{124}$Sn \cite{Allmond} & 50 & [10$\cdots$14] & J=9/2 & 5 \\\hline
			17.4 & 25.6 & 0.049 & 6.5 & 6.25 & $^{126}$Sn \cite{Allmond} & 50 & [24$\cdots$28] & J=9/2 & 4 \\\hline
			15.12 & 22.57 & 0.045 & 5.5 & 4.22 & $^{98}$Ru \cite{Radeck} & 44 & [15$\cdots$21] & J=9/2 & 4 \\\hline
			21.22 & 30.37 & 0.062 & 8.8 & 8.12 & $^{132}$Te \cite{Stone01} & 52 & [28$\cdots$32] & J=9/2 & 4 \\\hline
		\end{tabular}
		\label{table2}
	\end{center}
\end{table}

\begin{figure}[h!]
	\begin{center}\includegraphics[width=0.90\linewidth]{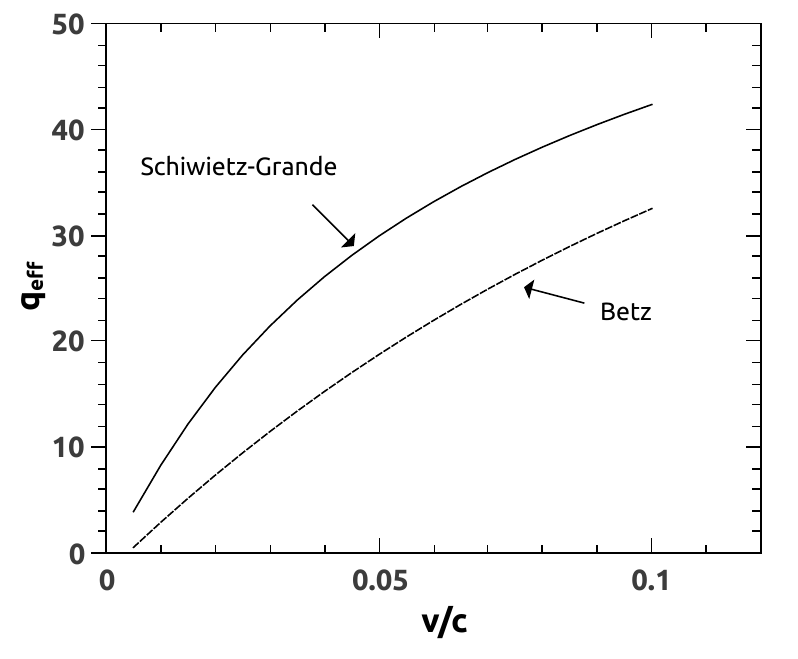}\end{center}
	\caption{Effetive charge $q_{eff}$ versus velocity for $^{148}$Nd and $Z_t = 6$.}
	\label{Betz_Grand}
\end{figure}

The observed linear dependence of the magnetic field (${B}$) on the effective charge divided by the major quantum number ($q_{eff}/n$) is consistent with the theoretical descriptions of ionization processes proposed by Bohr \cite{Bohr}, as well as the works of Northcliffe \cite{Nothcliffe} and Northcliffe and Schilling \cite{Noth01}. In 1972, Betz \cite{Betz} derived a semi-empirical relation for average charge states, which was later extended to cover a wide range of ion projectiles and targets from Be to Bi by Grande \cite{Grande}. Fig. \ref{Betz_Grand} compares the results obtained from both methods.

The data points presented in Fig. \ref{comp_qeff} were obtained using the formulas provided by Grande \cite{Grande}. The experimental points align with the calculations performed using the GRASP code \cite{Grasp,Grasp1,Grasp2}. In the experiments, available information is limited to the velocity of ions, but it gives the number of ionized electrons and the charge state. 

Another constraint on the electron configuration set was measured mean spin of the electron cloud, which, in almost all cases, was $J = 9/2$.\\

\vfill

\section{Summary}

The magnetic properties of atoms and nuclei were analyzed based on experimental data measuring the magnetic moment of the nucleus by the RIV method. The linear dependence of the magnetic field at the nucleus versus effective charge divided by the major quantum number was found and analyzed. This systematic dependence can be attributed to two main factors: first, the effective charge distribution along velocity of ions, and second, the total spin of electron cloud. This two boundaries significantly reduces the range of possible set of electron configurations.

It is also interesting that experimental measurements of the magnetic strength field at nucleus indicates that during the ionization process, electrons prefer to stay in pairs. There is less probable that one electron stay alone on $ns$ orbital. If that happen it rather prefer to de-excite on the lower orbital $d$.  

It is anticipated that future studies on other nuclei will contribute to the development of a systematic understanding of magnetic properties across the entire range of elements.\\

{\textbf{Acknowledgment.}}

The paper was partly financed by the international project “PMW” of the Polish Minister of Science and Higher Education; active in the period 2022-2024; grant Nr 5237/GSI-FAIR/2022/0.

\vfill



\begin{thebibliography}{99}
	\bibitem{AS} A. Stuchbery, Nuclear Physics A \textbf{700}, 83 (2002).
	\bibitem{Goertzel} G.Goertzel, Phys.Rev. \textbf{70} 897 (1946).
	\bibitem{Abragam} A. Abragam and R.V. Pound, Phys. Rev. \textbf{92}, 943 (1953).
	\bibitem{Landau} L.D. Landau and E. M. Lifshitz, Qauntum Mechanics, Non-relativistic theory, (Pergamon Press, Oxford, ()1977).
	\bibitem{Grasp} P. Jonsson, J. Biero, T. Brage, J. Ekman, C. Froese Fischer, G. Gaigalas, M. Godefroid, I.P. Grant, J. Grumer, The Computational Atomic Structure Group (2015), downloaded from: http://ddwap.mah.se/tsjoek/compas/.
	\bibitem{Grasp1} P. Jonsson, G. Gaigalas, J. Biero, C. Froese Fischer, and I.P. Grant Comput. Phys. Commun. \textbf{184}, 2197 (2013). C. Naze, E. Gaidamauskas, G. Gaigalas, M. Godefroid, and P. Jonsson Comput. Phys. Commun. \textbf{184}, 2187 (2013).
	\bibitem{Grasp2} P. Jonsson, X. He, C. Froese Fischer, and I. P. Grant, Comput. Phys. Commun. \textbf{177}, 597 (2007).
	\bibitem{Grasp02} Froese Fischer, C., Brage, T., Jonsson, P.: Computational Atomic Structure. IoP, London (1997). Froese Fischer, C., Tachiev, G.: At. Data Nucl. Data Tables \textbf{87}, 1 (2004).		
	\bibitem{Bohr} N. Bohr, K. Dan. Vidensk. Selsk. Mat.-Fys. Medd \textbf{18}, 8 (1948).
	\bibitem{Grande} G. Schiwietz and P.L. Grande Nucl. Instr. Meth. in Phys. Res. B \textbf{175-177}, 125 (2001).
	\bibitem{Betz} H.D. Betz, Rev. Mod. Phys. vol.\textbf{44}, 3 (1972).
	\bibitem{Andrews} H.R.Andrews et al. Hyperfine Interactions \textbf{4}, 110 (1978).
	\bibitem{Stuchbery_rev} A. Stuchbery, Hyperfine Interact \textbf{220}, 29 (2013).	
	\bibitem{Stone01} A. E. Stuchbery and N. J. Stone Phys. Rev. C \textbf{76}, 034307 (2007).
	\bibitem{Chen} X. Chen et al. PRC \textbf{87}, 044305 (2013).
	\bibitem{Dafni} E. Dafni, J. Bendahan, C. Broude, G. Goldring, M. Hass,	E. Naim, M. H. Rafailovich, C. Chasman, O. C. Kistner, and	S. Vajda, Nucl. Phys. A \textbf{443}, 135 (1985).
	\bibitem{Cern} K Nakamura et al (Particle Data Group), J. Phys. G: Nucl. Part. Phys. \textbf{37} 075021 (2010).	
	\bibitem{Ward} D. Ward, R.L. Graham, J.S. Geiger, H.R. Andrews, S.H Sie, Nucl. Phys. A \textbf{193}, 479 (1972).
	\bibitem{AT} A. Tucholski et al.,  to be published.	
	\bibitem{Allmond} J. M. Allmond et al. Phys. Rev. C \textbf{87}, 054325 (2013).
	\bibitem{Radeck} D.Radeck et al., Phys. Rev. C \textbf{85}, 014301 (2012).
	\bibitem{Nothcliffe} C. Northcliffe, Annual Review of Nuclear Science Vol. \textbf{13}, 67 (1963).
	\bibitem{Noth01} L.C.Northcliffe and R.F.Schilling, Atomic Data and Nuclear Data Tables, vol \textbf{7}, Issue 3-4, 233 (1970).
\end{thebibliography}
\end{document}